# All-dielectric phase-change reconfigurable metasurface

Artemios Karvounis[1], Behrad Gholipour[1], Kevin F. MacDonald[1], and Nikolay I. Zheludev[1,2]

[1] Optoelectronics Research Centre & Centre for Photonic Metamaterials, University of Southampton, Southampton, SO17 1BJ, UK

[2] Centre for Disruptive Photonic Technologies, School of Physical and Mathematical Sciences & The Photonics Institute, Nanyang Technological University, Singapore 637371

We harness non-volatile, amorphous-crystalline transitions in the chalcogenide phase-change medium germanium antimony telluride (GST) to realize optically-switchable, all-dielectric metamaterials. Nanostructured, subwavelength-thickness films of GST present high-quality resonances that are spectrally shifted by laser-induced structural transitions, providing reflectivity and transmission switching contrast ratios of up to 5:1 (7 dB) at visible/near-infrared wavelengths selected by design.

From their emergence as a paradigm for engineering new passive electromagnetic properties such as negative refractive index or perfect absorption, metamaterial concepts have extended rapidly to include a wealth of dynamic - switchable, tunable, reconfigurable, and nonlinear optical functionalities, typically through the hybridization of plasmonic (noble metal) metamaterials/surfaces with active media.[1] Phase-change materials, including chalcogenides,[2-5] vanadium dioxide,[6-8] gallium,[9] and liquid crystals[10-12] have featured prominently in this evolution. We now show that the chalcogenides offer a uniquely flexible platform for the realization of non-volatile, optically-switchable all-dielectric metamaterials. Subwavelength-thickness germanium antimony telluride (GST) nano-grating metasurfaces (Fig. 1) provide high-quality ($Q \geq 20$) near-infrared resonances that can be spectrally shifted by optically-induced crystallization to deliver reflection and transmission switching contrast ratios up to 5:1 (7 dB).

To mitigate the substantial Ohmic losses encountered in plasmonic metamaterials at optical frequencies, which compromise many applications, while also improving manufacturing process practicality and compatibility with established (opto)electronic technologies, considerable effort has been devoted of late to the realization of 'all-dielectric' metamaterials, presenting resonances based upon the excitation of Mie as opposed to plasmonic (displacement as opposed to conduction current) modes in high-index, low-loss dielectric as opposed to noble metal nanostructures. A wide range of passive all-dielectric metasurface planar optical elements for steering, splitting, filtering, focusing and variously manipulating beams have been demonstrated, very typically using silicon for visible to near-IR wavelengths.[13-17] Active functionalities have been demonstrated on the basis of hybridization of a silicon metasurface with a liquid crystal,[18] two photon absorption on silicon metasurfaces[19,20] and nonlinear optomechanical reconfiguration in a free-standing silicon membrane metasurface.[21]

By virtue of their compositionally-controlled high-index, low-loss characteristics, which extend over a broad spectral range from the visible to long-wave infrared, and can moreover be reversibly switched (electrically or optically) in a non-volatile fashion, the chalcogenides (binary and ternary sulphides, selenides and tellurides) provide an exceptionally adaptable material base for the realization of optically reconfigurable meta-devices. Their phase-change properties – reversible transitions between amorphous and crystalline states with markedly different optical and electronic properties – have been utilized for decades in optical data storage and more recently in electronic phase-change RAM.[22] The crystalline-to-amorphous transition is a melt-quenching process initiated by a short (few ns or less), intense excitation that momentarily raises the local temperature above the melting point $T_m$; the amorphous-to-crystalline transition is an annealing process requiring a longer (sub-µs), lower intensity excitation to hold the material above its glass transition temperature $T_g$ (but below $T_m$) for a short time. The latter can also be achieved through an accumulation of sub-threshold (including fs laser pulse) excitations, facilitating reproducible 'greyscale' and neuromorphic switching modes of interest for all-optical data and image processing, and harnessed recently for direct, reversible laser writing of planar optical elements and short/mid-wave IR metamaterials in a chalcogenide thin film.[23-26]

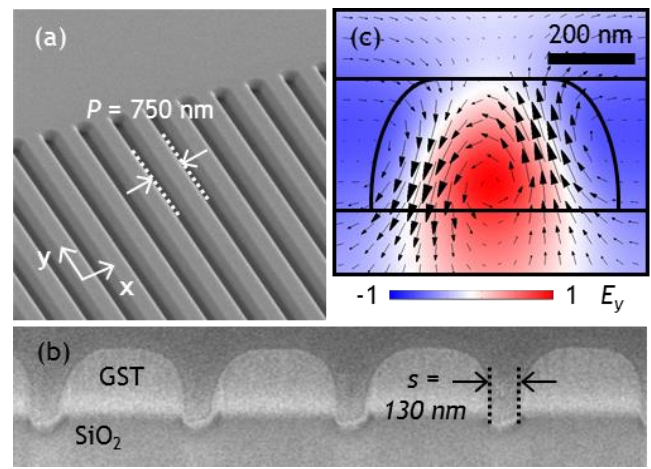

**Figure 1**. All-chalcogenide nano-grating metasurface. (a) Oblique incidence and (b) cross-sectional scanning electron microscope images of a 750 nm period grating fabricated by focused ion beam milling in a 300 nm thick amorphous GST film on silica. (c) Numerically simulated distribution of the y-component of electric field in the xz plane, overlaid with arrows denoting the direction and magnitude of magnetic field, for a unit cell of such a grating at resonance [at wavelength $\lambda = 1235$ nm for $P = 750$ nm; slot width $s = 130$ nm].

Here, we demonstrate structurally engineered high-quality near-infrared transmission and reflection resonances in planar (300 nm thick) dielectric nano-grating metasurfaces of amorphous germanium antimony telluride ($Ge_2Sb_2Te_5$ or GST - a widely used composition in data storage applications), and the non-volatile switching of these resonances via laser-induced crystallization of the chalcogenide. We employ nano-grating array metasurface patterns of subwavelength periodicity (Fig. 1), similar to those used, for example, in demonstrations of active nanophotonic photodetectors and tunable filters.[27-29] A thin (subwavelength) film of a transparent medium at normal incidence has properties of reflection and transmission dependent on its thickness and complex refractive index. Periodically structuring such a film on the subwavelength scale has the effect of introducing narrow reflection/transmission resonances via the interaction between thin film interference and grating mode.[30, 31] Such structures are non-diffractive and thus behave in the far field as homogenous layers.[32] In the case of anisotropic structuring, the resultant optical properties are dependent on the polarization of incident light.

GST films with a thickness of 300 nm were deposited on optically flat quartz substrates by RF sputtering (Kurt J. Lesker Nano 38). A base pressure of $5\times10^{-5}$ mbar is achieved prior to deposition and high-purity argon is used as the sputtering gas (70 ccpm to strike, 37 ccpm to maintain the plasma). The substrate is held within 10K of room temperature on a rotating platen 150 mm from the target to produce low-stress amorphous films. Nano-grating metasurface patterns, with a fixed slot width $s$ ~130 nm and periods $P$ from 750 to 950 nm, each covering an area of approximately 20 μm × 20 μm, were etched through the GST layer by focused ion beam (FIB) milling (Fig. 1). The normal-incidence transmission and reflection characteristics of these GST nano-grating metasurfaces were subsequently quantified, for incident polarizations parallel and perpendicular to the grating lines (along the $y$ and $x$ directions defined in Fig. 1, or TE and TM orientations of the grating, respectively), using a microspectrophotometer (CRAIC QDI2010) with a sampling domain size of 15 μm × 15 μm.

Unstructured, amorphous GST is broadly transparent in the near-IR range, with measured transmission at a thickness of 300 nm >70% between 1300 and 1800 nm (Fig. 2); absorption being <20% in this spectral range (<1% above 1500 nm). Nano-grating metasurface structures introduce pronounced resonances, with quality factors $Q$ more than 20 ($Q = \lambda_r/\Delta\lambda$ where $\lambda_r$ is the resonance frequency and $\Delta\lambda$ is the half-maximum linewidth), for TE polarized light, as shown in Figs. 2a and 2b, at spectral positions directly proportional to the nano-grating period $P$.

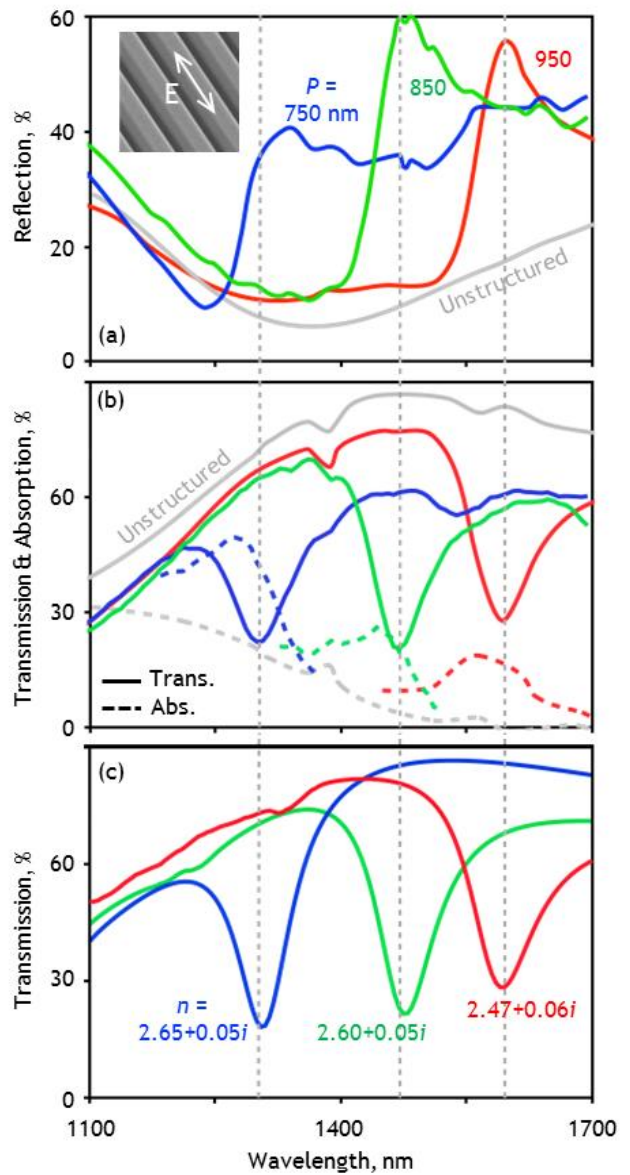

**Figure 2.** (a, b) Microspectrophotometrically measured Reflection $R$, transmission $T$ and absorption $A$ [$=1-\{R+T\}$] spectra for 300 nm thick amorphous GST nano-grating metamaterials with a selection of periods $P$ [as labelled; slot width $s$ = 130 nm], under TE-polarized illumination, alongside spectra for the unstructured amorphous GST film. (c) Numerically simulated transmission spectra calculated using a non-dispersive GST refractive index value for each grating period [as labelled] selected to reproduce experimental resonance positions and widths.

This behavior is replicated in finite element (Comsol MultiPhysics) numerical simulations. These assume in all cases a lossless non-dispersive refractive index of 1.46 for the silica substrate, normally incident narrowband plane wave illumination and, by virtue of periodic boundary conditions, a grating pattern of infinite extent in the $xy$ plane. Using in the first instance ellipsometrically obtained values for the complex refractive index of the unstructured amorphous GST film (see Supplemental Material[33]), which has a weakly dispersive real part ~2.5 (decreasing slowly with increasing wavelength from 2.6 to 2.4 between 1000 and 1700 nm) and a loss coefficient <0.045 across most of the near-IR spectral range (rising below 1130 nm to reach 0.09 at 1000 nm), a good qualitative match is obtained, albeit with values of $Q > 30$ (see Supplemental Fig. S2 [33]). Using instead the real and imaginary parts of GST refractive index as free parameters in the model, an improved match to the experimentally observed spectral positions and widths of nano-grating resonances is achieved with non-dispersive refractive index values for each grating period (as shown in Fig. 2c) that are marginally higher in both the real and imaginary part than the ellipsometric values at the corresponding resonant wavelength. The discrepancies between ellipsometrically derived and experimental/fitted spectra are related to manufacturing imperfections, i.e. deviations from the ideal model geometry such as slight over-milling of grating lines into the substrate, and to contamination / stoichiometric change in the GST during FIB milling, which effects some change in refractive index.

The simulations show that the TE resonance is associated with the excitation of anti-phased displacement currents (in the $\pm y$ direction) along the core and sides of each GST 'nanowire', and a circulating pattern of magnetic field centered within the wire as illustrated in Fig. 1c. For the orthogonal TM polarization, the spectral dispersion of reflectivity and transmission (as measured by the microspectrophotometer, which employs an objective of numerical aperture 0.28) is more complex: Resonances, again at wavelengths proportional to $P$, are split as a result of the structures' sensitivity in this orientation to the incident angle of light (see Supplemental Material[33]).

By engaging the phase-change properties of GST,[22] the resonances of all-chalcogenide metasurfaces can be optically switched in a non-volatile fashion. In the present case, GST nano-gratings are converted from the as-deposited amorphous phase to a crystalline state by laser excitation at a wavelength of 532 nm (selected for its strong absorption in GST). This annealing is achieved by raster-scanning the beam, with a spot diameter of ~5 μm and continuous wave intensity of ~3 mW/μm$^2$, over the sample to bring the GST momentarily to a temperature above its glass-transition point $T_g$ but below its melting point $T_m$ (around 110 and 630°C respectively,[34, 35] though exact values will vary with factors including film thickness, composition and density).

The resultant change in GST's complex refractive index produces a change in the spectral dispersion of the nano-grating resonances, bringing about substantial changes in the metasurface transmission and reflection, especially at wavelengths close to the resonance – absolute levels are seen to increase/decrease by as much as a factor of five (Fig. 3). An increase in the real part of the GST refractive index red-shifts metasurface resonances by approximately 150 nm, while the concomitant increase in the imaginary part of the index is primarily responsible for increasing the resonance linewidth and broadband (non-resonant) absorption, particularly at shorter wavelengths.

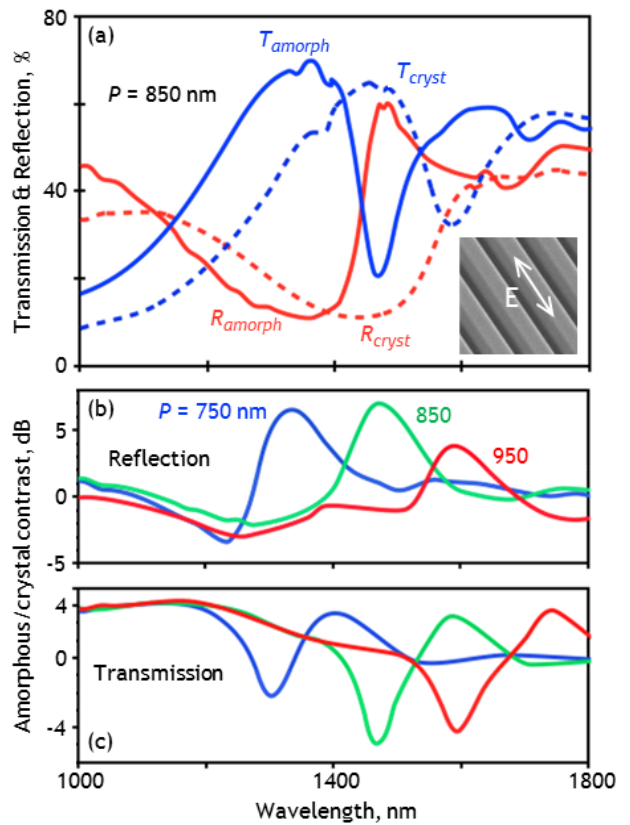

**Figure 3.** (a) Microspectrophotometrically measured TE-mode reflectivity and transmission spectra for the as-deposited amorphous and laser-annealed [partially] crystalline phases of a 300 nm thick GST nano-grating metamaterial with a period $P$ = 850 nm [slot width $s$ = 130 nm]. (b, c) Spectral dispersion of TE-mode reflection (b) and transmission (c) switching contrast, evaluated as $10 log(A/C)$ where $A$ and $C$ are respectively the amorphous and crystalline levels, for a selection of GST nano-grating periods [as labelled].

Repeating the above process of matching numerically modelled spectra to experimentally observed nano-grating resonance positions and widths yields (for all three grating periods, again under the non-dispersive approximation) a refractive index value of $2.85 + 0.09i$ for the laser-annealed GST in the nano-gratings (see Supplemental Fig. S3 [33]). The induced index change of ~10% is substantial but yet smaller and less wavelength-dependent than may be expected on the basis of ellipsometric data for unstructured crystalline GST[33]. This indicates strongly that the nanostructured GST is stoichiometrically modified and/or only partially crystallized,[5, 36, 37] which is to be anticipated primarily as a consequence of the FIB milling process (reduction of refractive index due to irradiation,[38] creation of defects and gallium implantation) and because nanostructuring unavoidably modifies the thermal properties of the film (i.e. the energy absorbed from the laser beam at a given point; the temperature achieved; and the rates of temperature increase/decrease).

Nonetheless, the refractive index change (resonance spectral shift) achieved experimentally may be considered close to optimal in that it maximizes transmission and reflectivity contrast. For example, the 1470 nm reflectivity maximum of the amorphous 850 nm period grating becomes a reflectivity minimum in the (partially) crystalline state, giving a contrast ratio ($R_{amorphous}$:$R_{crystalline}$) of 5:1 (7 dB).

The reverse crystalline-to-amorphous transition is not demonstrated as part of the present study: This melt-quench process would require transient heating of the GST to a temperature above $T_m$, which would lead to geometric deformation and chemical degradation of the samples. Robust metasurfaces supporting reversible switching over many cycles may be realized by encapsulating the GST nanostructure in the manner of the functional chalcogenide layers within rewritable optical discs, which are located between protective layers of $ZnS$:$SiO_2$. In this regard, all-chalcogenide metasurfaces hold a notable advantage over hybrid plasmonic-metal/chalcogenide metamaterials,[2-5] which ultimately require similar passivation layers (and thereby inevitably sacrifice switching contrast due to the separation necessarily introduced between the active chalcogenide component and the surface, i.e. optical near-field, of the plasmonic metal). All-chalcogenide metasurfaces can also offer lower insertion losses and greater ease of fabrication (via a single lithographic step in a single layer of CMOS-compatible material) than plasmonic hybrid structures.

In summary, we have realized all-dielectric photonic metasurfaces using a chalcogenide phase-change material platform, and demonstrated high-contrast, non-volatile, optically-induced switching of their near-infrared resonant reflectivity and transmission characteristics. Subwavelength (300 nm; <$\lambda$/5) films of germanium antimony telluride (GST) structured with non-diffractive sub-wavelength grating patterns, present high-quality resonances that are spectrally shifted by as much as 10% as a consequence of a laser-induced (amorphous-crystalline) structural transitions in the GST, providing switching contrast ratios of up to 5:1 (7 dB) in reflection and 1:3 (-5 dB) in transmission.

Within the transparency range of the (unstructured) host medium, high switching contrast wavebands can be engineered by design, i.e. appropriate selection of metasurface pattern geometry and dimensions. GST – a material with an established industrial footprint in optical and electronic data storage, can readily be structured for telecommunications applications at 1550 nm, while other members of the extensive (sulphide, selenide and telluride) chalcogenide family may also provide similar active, all-dielectric metasurface functionality in the visible range and at infrared wavelengths out to 20 μm.

Among material platforms for all-dielectric metamaterials, chalcogenides offer unparalleled compositional variety (i.e. range and variability of material parameters) and non-volatile (including binary[22] as well as incremental 'greyscale'[23]) switching functionality. A wealth of reconfigurable and self-adaptive subwavelength-thickness 'flat-optic' applications may be envisaged, including switchable/tunable bandpass filter, lens, beam deflection and optical limiting components. It is interesting to note that the GST metasurfaces' resonant reflectivity and transmission changes occur in the opposite direction to those in the unstructured chalcogenide (e.g. crystallization increases the reflectivity of an unstructured film but decreases that of a nano-grating at resonance). The ability of a single medium to provide both high and low reflectivity/transmission (signal on and off) levels in the same phase state, such that they can be simultaneously inverted via a homogenous, sample-wide structural transition, may be of interest in image processing as well as the above metasurface optics applications.

This work was supported by the Engineering and Physical Sciences Research Council [grants EP/G060363/1 and EP/M009122/1], the Samsung Advanced Institute of Technology [collaboration project IO140325-01462-01], The Royal Society, the Singapore Ministry of Education [grant MOE2011-T3-1-005], and the Singapore Agency for Science, Technology and Research (A*STAR) [SERC project 1223600007]. Following a period of embargo, the data from this paper will be available from the University of Southampton repository at http://dx.doi.org/10.5258/SOTON/392918.

# All-dielectric phase-change reconfigurable metasurface: Supplemental Material

Artemios Karvounis[1], Behrad Gholipour[1], Kevin F. MacDonald[1], and Nikolay I. Zheludev[1, 2]

[1] Optoelectronics Research Centre & Centre for Photonic Metamaterials, University of Southampton, Southampton, SO17 1BJ, UK

[2] Centre for Disruptive Photonic Technologies, School of Physical and Mathematical Sciences & The Photonics Institute, Nanyang Technological University, Singapore 637371

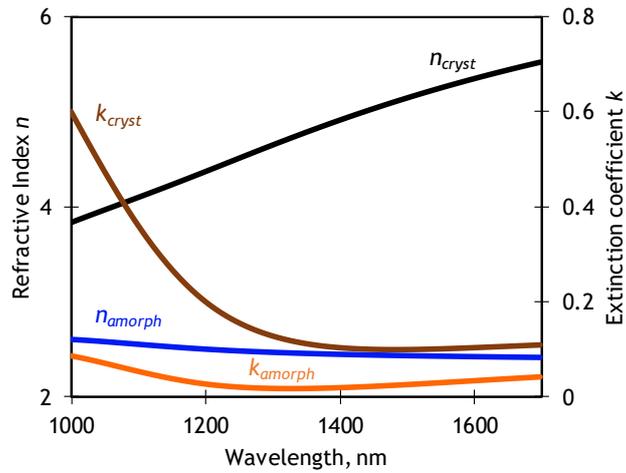

**Figure S1**. Near-IR dispersion of complex refractive indices, from spectroscopic ellipsometry, of the as-deposited amorphous and laser-annealed crystalline phases of a 300 nm thick unstructured Ge:Sb:Te film on silica.

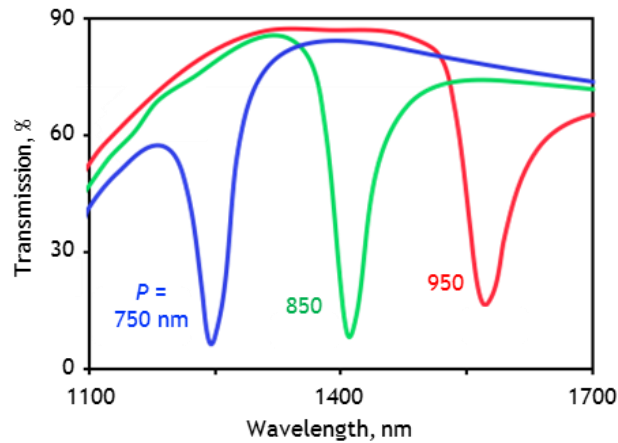

**Figure S2**. Numerically simulated transmission spectra for 300 nm thick amorphous GST nano-grating metamaterials with a selection of periods [as labelled; slot width $s$ = 130 nm], under TE-polarized illumination, calculated using ellipsometircally measured values for the [weakly dispersive] complex refractive index of unstructured amorphous GST.

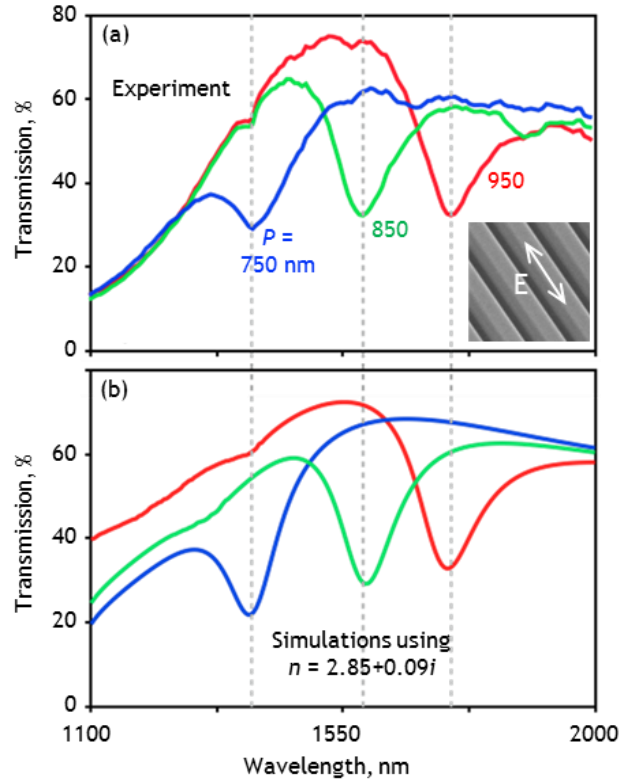

**Figure S3.** (a) Microspectrophotometrically measured transmission spectra for 300 nm thick laser-annealed [partially] crystalline GST nano-grating metamaterials with a selection of periods $P$ [as labelled; slot width $s$ = 130 nm], under TE-polarized illumination. (b) Corresponding numerically simulated transmission spectra calculated using a non-dispersive GST refractive index value of $2.85 + 0.09i$ that reproduces the experimental resonance positions and widths for all grating periods.

----------------------

**TM-polarized illumination**

Under TM-polarized illumination nano-grating resonances are split (Supplementary Fig. S4) as a result of the structures' strong sensitivity in this orientation to the incident angle of light: The microspectrophotometer employs an objective with a numerical aperture of 0.28, thereby illuminating samples at incident angles $\theta$ ranging from zero and ~16°. This is of little consequence to the TE mode (experimental data are reproduced well by numerical simulations assuming ideally normal incidence), but for the TM polarization spatial symmetry is broken by the slightest deviation from normal incidence, leading to the observed resonance splitting.[1,2] The TM resonances are characterized by displacement currents circulating in the *xz* plane (forming magnetic dipoles oriented along *y*) – a single symmetric loop centered within each nanowire at singularly normal incidence, $\theta = 0°$; more complex asymmetric double-loop distributions at off-normal angles (Supplementary Fig. S5).

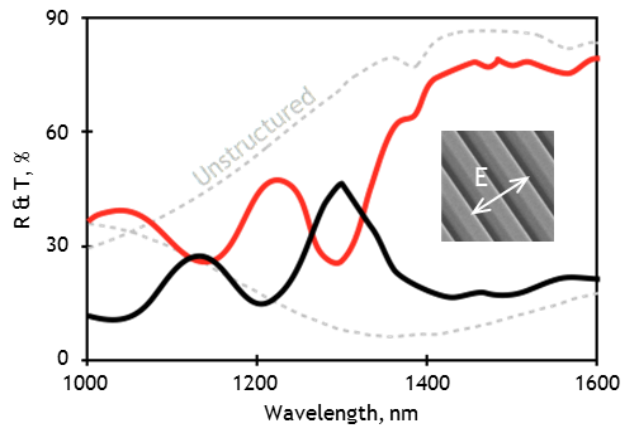

**Figure S4.** Microspectrophotometrically measured reflection and transmission spectra for a 300 nm thick amorphous GST nano-grating metamaterial with a period $P$ = 750 nm [slot width $s$ = 130 nm], under TM-polarized illumination.

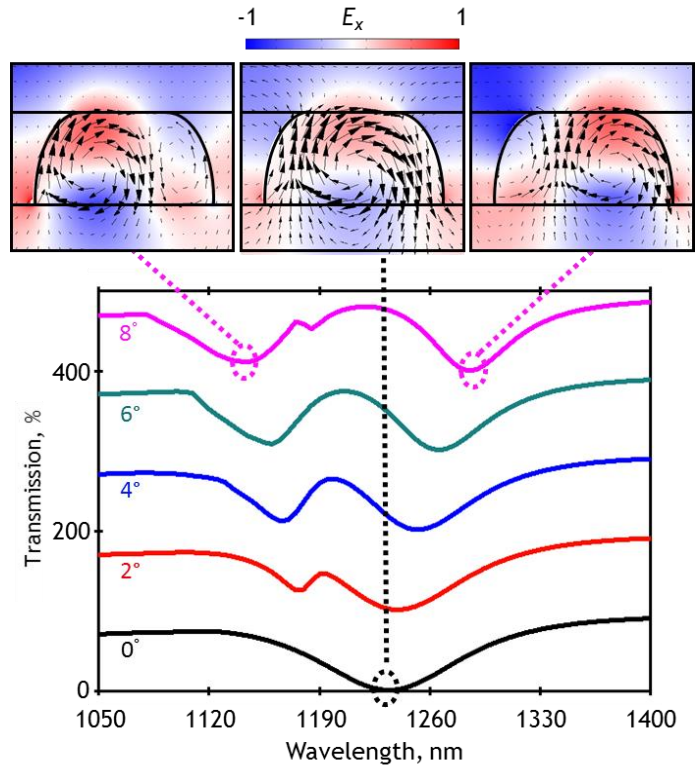

**Figure S5**. Numerically simulated TM-mode transmission spectra for a 300 nm thick GST nano-grating metamaterial, with a period $P = 750$ nm and slot width at the lower and upper surfaces of the GST layer of 130 and 450 nm respectively, for incident angles between 0° and 8° [as labelled; vertically offset for clarity]. Field maps above show the distribution of the x-component of electric field in the *xz* plane for a unit cell of the metasurface at the singular normal incidence resonance [$\lambda = 1235$ nm] and the two minima [= 1145 and 1285 nm] of the split resonance for an incident angle of 8°.